\documentclass{article}

\usepackage{PRIMEarxiv}
\usepackage[utf8]{inputenc} % allow utf-8 input
\usepackage[T1]{fontenc}    % use 8-bit T1 fonts
\usepackage{hyperref}       % hyperlinks
\usepackage{url}            % simple URL typesetting
\usepackage{booktabs}       % professional-quality tables
\usepackage{amsfonts}       % blackboard math symbols
\usepackage{nicefrac}       % compact symbols for 1/2, etc.
\usepackage{microtype}      % microtypography
\usepackage{lipsum}
\usepackage{fancyhdr}       % header
\usepackage{graphicx}       % graphics
\graphicspath{{media/}}     % organize images and figures under media/ folder
\usepackage{amsmath}        % for math formatting
\usepackage{algorithm}      % for algorithm float environment
\usepackage{algorithmic}    % for pseudocode formatting
\usepackage{tikz}           % for drawing graphics
\usetikzlibrary{arrows, positioning}  % for arrow and node positioning in TikZ

\title{An Autonomous RL Agent Methodology for Dynamic Web UI Testing in a BDD Framework}

\author{ \href{https://orcid.org/0000-0002-0724-9197}{\includegraphics[scale=0.06]{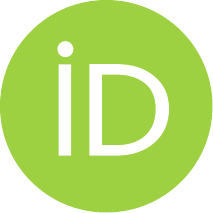}\hspace{1mm}Ali Hassaan Mughal} \\
        MSc. Computer Science\\
	Kansas State University\\
}

\begin{document}
\maketitle

\begin{abstract}
Modern software applications demand efficient and reliable testing methodologies to ensure robust user interface functionality. This paper introduces an autonomous reinforcement learning (RL) agent integrated within a Behavior-Driven Development (BDD) framework to enhance UI testing. By leveraging the adaptive decision-making capabilities of RL, the proposed approach dynamically generates and refines test scenarios aligned with specific business expectations and actual user behavior. A novel system architecture is presented, detailing the state representation, action space, and reward mechanisms that guide the autonomous exploration of UI states. Experimental evaluations on open-source web applications demonstrate significant improvements in defect detection, test coverage, and a reduction in manual testing efforts. This study establishes a foundation for integrating advanced RL techniques with BDD practices, aiming to transform software quality assurance and streamline continuous testing processes.
\end{abstract}

\section*{Problem Statement}

Modern web applications are characterized by their dynamic and intricate navigational structures, where each page and user interaction forms part of a complex network. Traditional testing methodologies often fall short in fully exploring these complex pathways, especially when user actions are interdependent and context-sensitive. This research proposes the integration of an autonomous reinforcement learning (RL) agent within a Behavior-Driven Development (BDD) framework to automate and enhance web UI testing by modeling the website as a maze.

The core idea is to define constant starting points---such as a homepage, login page, or any predefined entry condition---and distinct endpoints that signal the completion of a user scenario. Endpoints are identified by cues such as specific text presence, visible UI elements, or states where no further actions are possible. For example, in the context of an e-commerce platform like Amazon, the agent would:
\begin{itemize}
    \item Begin at a fixed starting point (e.g., the Amazon homepage or sign-in page).
    \item Navigate through the website by detecting and interacting with UI elements, such as searching for a product, adding it to the cart, and proceeding to checkout.
    \item Recognize endpoints such as a successful login confirmation, product detail verification, or an order confirmation page.
\end{itemize}

An \textbf{input summary} is provided to the agent, specifying the functionality to be tested and establishing the starting point. For example, the input summary may define a scenario like "Create an order for a specific product type" or "Post a review for a selected product category." The RL agent learns to maneuver between these defined points, autonomously exploring multiple pathways and uncovering various possibilities to reach the desired end state.

Key aspects of the problem include:
\begin{enumerate}
    \item \textbf{Modeling the Website as a Maze:} Treating each webpage as a node and each user interaction as a transition captures the inherent complexity of web navigation.
    \item \textbf{Defining Constant Starting Points:} Establishing consistent entry points such as the homepage, login page, or other specific landing pages.
    \item \textbf{Identifying Endpoints:} Clearly defining termination points based on cues such as confirmation messages, error notifications, or terminal states.
    \item \textbf{Input Summary for Test Scenarios:} Using a concise summary to indicate the targeted functionality, which allows the RL agent to concentrate on achieving the scenario-specific goal.
    \item \textbf{Learning and Exploration:} Allowing the RL agent to learn optimal pathways through interactions with the UI, thereby uncovering multiple valid routes from the starting point to the endpoint, and capturing diverse test scenarios.
\end{enumerate}

This framework not only facilitates the generation and execution of test cases but also adapts dynamically to the evolving structure of modern web applications. The integration of RL within a BDD context ensures that the resulting tests are both comprehensive and interpretable, bridging the gap between automated testing efficiency and human-understandable test scenarios.

\section{Introduction}

Modern software applications demand robust, scalable, and efficient testing methodologies to guarantee high-quality user interface functionality. Traditional manual and scripted testing approaches have proven insufficient in the face of dynamic web environments, where user interactions resemble navigating through a complex maze. Recent advances in reinforcement learning (RL) and Behavior-Driven Development (BDD) offer promising alternatives. This paper presents an autonomous testing agent that leverages RL and integrates with a BDD framework to generate, execute, and refine test scenarios aligned with specific business requirements and real user behaviors.

The concept of modeling websites as mazes, with fixed starting points (e.g., homepage, login page) and well-defined endpoints (e.g., confirmation pages, error states), is central to this approach. In this system, an \emph{input summary} describing the target functionality (e.g., "create an order" or "post a review") serves as the guide for the RL agent. The agent learns to explore multiple routes between these points, identifying valid transitions and capturing diverse test scenarios. This method is intended to enhance test coverage, increase defect detection, and reduce the need for manual test script maintenance.

Autonomous agents have been explored in various domains. Early environments for testing and automation \cite{1,27} laid the groundwork for realistic web interaction. Subsequent works extended these ideas by incorporating RL techniques to overcome the limitations of static and scripted test cases \cite{2,3,28}. In parallel, studies focusing on HTML understanding and multimodal perception have demonstrated the potential of large language models (LLMs) and visual language models in parsing complex web pages and generating actionable insights \cite{8,9,6}. The integration of these technologies is crucial for building agents capable of seamless interaction with diverse and evolving web interfaces.

Recent advances in autonomous agent design have shown that incorporating elements such as state-space exploration, reasoning traces, and dynamic feedback can significantly improve performance on real-world tasks \cite{10,16,20,21,26}. The present work draws on these developments to integrate an RL agent that not only explores web environments but also learns to follow BDD scenarios, thereby bridging the gap between automated execution and human-readable test cases.

\section{Literature Review}

The literature on autonomous web agents, RL-based testing, and BDD integration spans several interconnected domains. This section reviews the work in three primary categories: realistic web environments and datasets, RL and LLM-based autonomous agents, and BDD/automated UI testing.

\subsection{Realistic Web Environments and Datasets}

Several studies have focused on constructing realistic environments for training and evaluating web agents. For instance, \cite{1} introduced \emph{WebArena}, a comprehensive environment that emulates real-world websites across multiple domains, thereby addressing the shortcomings of synthetic testbeds. Similarly, \cite{2} presented \emph{Mind2Web}, a dataset spanning numerous websites and tasks, specifically designed for evaluating generalist web agents. Earlier platforms such as \emph{World of Bits} \cite{27} laid the foundational concepts by representing web pages as complex, interactive environments that serve as benchmarks for RL agents. These works collectively underscore the need for environments that capture the dynamic and intricate nature of real-world web interfaces.

\subsection{RL and LLM-based Autonomous Agents}

Reinforcement learning is a pivotal technique for developing autonomous agents for sequential decision-making tasks. For example, \cite{3} explored the use of RL for generating UI actions based on pixel inputs, demonstrating that agents can learn effective strategies by mimicking human interactions. In the domain of computer control, \cite{4} provided large-scale datasets for Android device control, emphasizing the importance of multi-step tasks that require both semantic and visual understanding.

More recent studies have focused on integrating LLMs into autonomous agent architectures. \cite{5} proposed the ReWOO framework to decouple reasoning from observations, significantly improving token efficiency during test case generation. The \emph{CogAgent} model \cite{6} demonstrated the capability of visual language models to comprehend and navigate graphical user interfaces, which is critical for realistic UI testing. Additionally, \cite{7} introduced Synatra, an approach that converts indirect knowledge (e.g., online tutorials) into direct demonstrations, enabling scalable training for digital agents.

Advancements in multimodal understanding have been particularly influential. Research on instruction-finetuned models for multimodal web navigation \cite{8} and HTML understanding \cite{9} has shown that combining textual and visual inputs leads to more robust agent performance. Autonomous agents such as those developed in \cite{10,11,12} integrate both visual and textual modalities to interact effectively with real-world websites. The integration of planning and long-context understanding, as demonstrated by \cite{13} and \cite{14}, further underscores the importance of combining various learning paradigms to achieve high performance in complex tasks.

Recent studies, including \cite{15} (AgentBench) and \cite{16} (Ferret-UI), have benchmarked the performance of LLM-based agents, highlighting the gap between human-level performance and current state-of-the-art models in diverse testing scenarios. Techniques emphasizing executable actions and dynamic feedback \cite{17,18} provide further insights into improving agent performance by integrating recursive critique and refinement strategies. Moreover, \cite{19} highlights the importance of harmonizing natural language and code to create agents capable of both reasoning and execution.

\subsection{BDD and Automated UI Testing}

Behavior-Driven Development (BDD) has become an industry-standard practice to ensure that software behavior aligns with user expectations. The integration of RL with BDD, as proposed in this paper, leverages the strengths of both paradigms. Traditional automated UI testing often relies on scripted or rule-based approaches, which lack adaptability in dynamic web environments. In contrast, the proposed framework allows an RL agent to autonomously generate BDD scenarios by learning directly from web interactions.

Efforts to bridge the gap between automated testing and human-readable specifications are evident in related research. For example, studies on dynamic scenario reusability and step auto-completion for frameworks like Cucumber \cite{31} demonstrate the need for intelligent agents that integrate smoothly with existing BDD workflows. Research on automated test generation \cite{20,21,22,23} further supports the theoretical and practical basis for the proposed system.

\subsection{Synthesis and Research Gap}

The reviewed literature indicates a clear trend toward leveraging advanced RL techniques, multimodal LLMs, and realistic web environments for building autonomous agents. However, significant challenges persist in integrating these technologies within a unified framework for UI testing. Although individual components—such as realistic environment simulation \cite{1,2,27}, RL-based agent learning \cite{3,4,5,7,10,14}, and multimodal understanding \cite{6,8,9,12}—have been extensively studied, synthesizing these approaches within a BDD context remains an open research question.

This paper addresses this gap by proposing a framework that integrates an autonomous RL agent with BDD practices for web UI testing. The aim is not only to improve test coverage and defect detection rates but also to generate human-interpretable test cases that are easily incorporated into continuous integration pipelines.

\section{Implementation Details}

This section describes the implementation of the autonomous RL agent integrated within a BDD framework for web UI testing. The system is designed to automatically generate and refine test scenarios by modeling the website as a maze. The system accepts an input summary that defines the testing functionality (e.g., "Create an order for a specific product") and navigates from predefined starting points (e.g., homepage or login page) to designated endpoints (e.g., order confirmation).

\subsection{Scenario: Creating an Order on an E-Commerce Platform}

Consider an e-commerce scenario:
\begin{quote}
    \textbf{Input Summary:} "Place an order for a specific product category (e.g., electronics)."
\end{quote}
In this scenario, the RL agent is responsible for:
\begin{enumerate}
    \item \textbf{Initialization:} Loading the homepage and, if required, performing user authentication.
    \item \textbf{Exploration:} Navigating through the website by accessing menus, filtering by the electronics category, and searching for products.
    \item \textbf{Action Execution:} Selecting a product, adding it to the shopping cart, proceeding to checkout, and simulating a payment process.
    \item \textbf{Endpoint Recognition:} Detecting the order confirmation page by identifying cues such as confirmation messages or tracking numbers.
    \item \textbf{Learning and Refinement:} Updating the policy based on received rewards and refining the exploration strategy using backtracking when dead-ends are encountered.
\end{enumerate}

\subsection{System Architecture and Algorithmic Components}

The system is composed of the following modules:
\begin{itemize}
    \item \textbf{Input Parser:} Extracts key functionalities from the input summary and defines the target scenario.
    \item \textbf{State Representation Module:} Encodes the current webpage state using a combination of DOM extraction, visual recognition (via convolutional neural networks or transformer-based vision encoders), and text processing.
    \item \textbf{Action Space Definition:} Defines a set of generic actions such as \texttt{click(element)}, \texttt{type(text, element)}, and \texttt{scroll(direction)}. The choice of a discrete set of actions is driven by the need for clarity in execution and ease of mapping to UI events.
    \item \textbf{Reward Mechanism:} Provides intermediate rewards for detecting relevant cues (e.g., product details, cart updates) and final rewards upon reaching an endpoint. The design of the reward function is critical and is tuned to balance exploration and exploitation.
    \item \textbf{Exploration and Learning Engine:} Utilizes RL algorithms to explore the website. In this system, the following algorithms are employed:
    \begin{itemize}
        \item \textbf{Deep Q-Networks (DQN):} This method is suitable for environments with discrete action spaces. DQN approximates the optimal Q-value function, guiding the agent to select actions that maximize cumulative rewards. The algorithm was chosen for its relative simplicity and effectiveness in discrete control tasks.
        \item \textbf{Policy Gradient Methods:} Techniques such as REINFORCE and Actor-Critic are used for scenarios with larger or continuous action spaces. These methods directly optimize the policy, which can be advantageous when the action space cannot be easily discretized. Their inclusion provides flexibility and improved performance in complex tasks.
        \item \textbf{Epsilon-Greedy Exploration:} This strategy is used to balance exploration and exploitation. The agent selects random actions with a probability $\epsilon$, which decays over time, allowing for more focused exploration as the policy converges.
        \item \textbf{Backtracking Mechanisms:} Inspired by dynamic programming, these mechanisms allow the agent to revisit earlier states and explore alternative paths when encountering dead-ends. This ensures a more comprehensive exploration of the UI.
    \end{itemize}
    \item \textbf{BDD Integration Module:} Transforms successful navigation trajectories into human-readable BDD scenarios (e.g., in Gherkin syntax), facilitating integration into continuous integration pipelines.
\end{itemize}

\subsection{Algorithmic Process}

The following pseudocode summarizes the exploration and learning process of the agent:

\begin{algorithm}[H]
\caption{Autonomous Web UI Testing Agent}
\label{alg:agent}
\begin{algorithmic}[1]
\REQUIRE Input summary $S$, starting state $s_0$, set of endpoints $E$, maximum iterations $N$
\STATE Initialize agent policy $\pi$ using DQN/Policy Gradient techniques
\STATE Initialize exploration parameter $\epsilon$ (for epsilon-greedy strategy)
\STATE Initialize memory $M \gets \emptyset$
\FOR{$i = 1$ to $N$}
    \STATE $s \gets s_0$ \COMMENT{Reset to starting point for each episode}
    \WHILE{$s \notin E$ \AND \textbf{not timeout}}
        \IF{random() $< \epsilon$}
            \STATE $a \gets \text{random action from action space}$
        \ELSE
            \STATE $a \gets \max_a Q(s,a)$ \COMMENT{Using DQN or sampling from $\pi(s, S)$}
        \ENDIF
        \STATE Execute action $a$, observe new state $s'$ and reward $r$
        \STATE Store transition $(s, a, r, s')$ in memory $M$
        \STATE $s \gets s'$
    \ENDWHILE
    \IF{$s \in E$}
        \STATE Update policy $\pi$ using the collected trajectory (via backpropagation in DQN or policy gradient update)
    \ELSE
        \STATE Apply backtracking: revisit previous states from $M$ and attempt alternative actions
    \ENDIF
    \STATE Optionally decay $\epsilon$ over time
\ENDFOR
\STATE Translate best-performing trajectories into BDD scenarios (e.g., in Gherkin syntax)
\end{algorithmic}
\end{algorithm}

\subsection{Graphical Workflow}

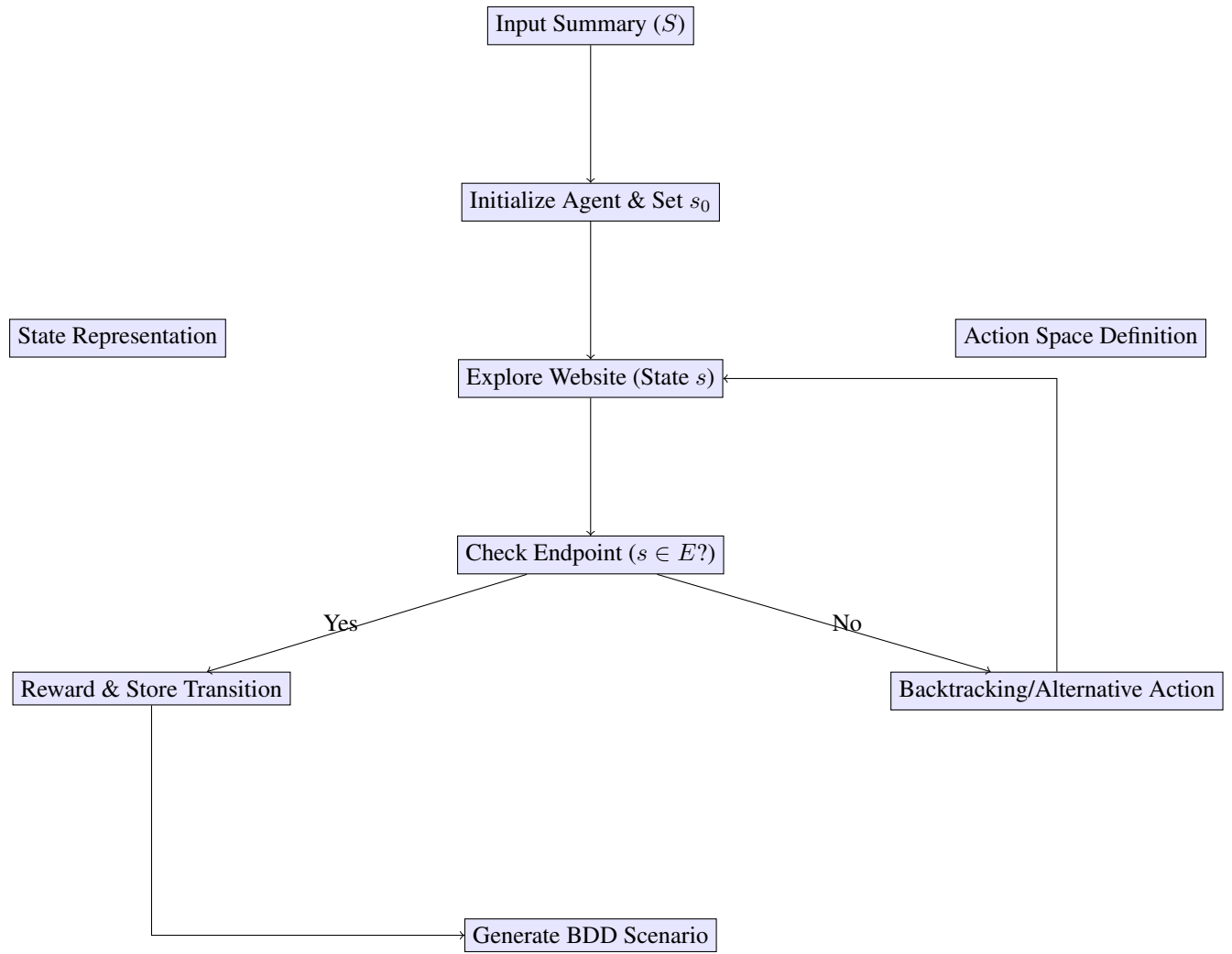
\begin{figure}[htbp]
    \centering
    \begin{tikzpicture}[node distance=2cm, auto]
        % Define nodes with proper horizontal spacing
        \node [draw, rectangle, fill=blue!10] (input) {Input Summary ($S$)};
        \node [draw, rectangle, fill=blue!10, below=of input] (init) {Initialize Agent \& Set $s_0$};
        \node [draw, rectangle, fill=blue!10, below left=of init, xshift=-2cm] (state) {State Representation};
        \node [draw, rectangle, fill=blue!10, below right=of init, xshift=2cm] (action) {Action Space Definition};
        \node [draw, rectangle, fill=blue!10, below=of init] (explore) {Explore Website (State $s$)};
        \node [draw, rectangle, fill=blue!10, below=of explore] (check) {Check Endpoint ($s \in E$?)};
        \node [draw, rectangle, fill=blue!10, below left=of check, xshift=-1cm] (reward) {Reward \& Store Transition};
        \node [draw, rectangle, fill=blue!10, below right=of check, xshift=1cm] (backtrack) {Backtracking/Alternative Action};
        \node [draw, rectangle, fill=blue!10, below=of check, yshift=-3cm] (bdd) {Generate BDD Scenario};
        
        % Draw arrows connecting the nodes
        \draw [->] (input) -- (init);
        \draw [->] (init) -- (explore);
        \draw [->] (explore) -- (check);
        \draw [->] (check) -- node[anchor=east] {Yes} (reward);
        \draw [->] (check) -- node[anchor=west] {No} (backtrack);
        \draw [->] (reward) |- (bdd);
        \draw [->] (backtrack) |- (explore);
    \end{tikzpicture}
    \caption{Workflow of the Autonomous Web UI Testing Agent. The agent receives an input summary, initializes with a defined state, explores the website using an RL-based policy (employing DQN, Policy Gradient methods, and epsilon-greedy exploration), checks for endpoint conditions, applies rewards or backtracking, and finally converts successful trajectories into BDD scenarios.}
    \label{fig:workflow}
\end{figure}

Figure~\ref{fig:workflow} illustrates the step-by-step process of the agent. Nodes are arranged with adequate horizontal and vertical spacing to ensure clarity.

\subsection{Discussion of Algorithmic Choices}

The selection of specific RL algorithms in this framework has been driven by their suitability for different aspects of the problem:
\begin{itemize}
    \item \textbf{Deep Q-Network (DQN):} DQN is particularly effective in environments where the action space is discrete. It approximates the optimal Q-value function, thereby guiding the agent to select actions that maximize cumulative rewards. The simplicity of DQN makes it a natural choice for tasks where UI interactions can be discretized.
    \item \textbf{Policy Gradient Methods:} In scenarios where the action space is large or continuous, methods like REINFORCE or Actor-Critic provide a direct optimization of the policy. These methods allow for more nuanced control in complex tasks and overcome the limitations associated with the discretization of actions. The Actor-Critic variant combines the benefits of both value-based and policy-based methods, providing more stable convergence.
    \item \textbf{Epsilon-Greedy Exploration:} This exploration strategy ensures that the agent does not prematurely converge to suboptimal policies by balancing random exploration and the exploitation of known good actions. The decay of $\epsilon$ over time is crucial for transitioning from exploration to exploitation.
    \item \textbf{Backtracking Mechanisms:} In web environments, dead-ends are common due to the dynamic nature of UIs. The backtracking mechanism enables the agent to recover from such states by revisiting earlier states and attempting alternative paths. This approach is inspired by dynamic programming techniques and enhances overall exploration efficiency.
\end{itemize}

This integrated system continuously refines its understanding of the website's structure, modeled as a maze, and learns optimal navigation paths from the starting points to the endpoints as defined by the input summary. Successful trajectories are transformed into human-readable BDD scenarios, bridging the gap between automated testing and manual verification.

\section{Results and Analysis}

This section outlines the potential outcomes and analysis methods to evaluate the performance of the proposed RL-driven BDD UI testing framework. The results can be visualized using various metrics and plots, providing a comprehensive understanding of the agent's performance in navigating and testing web applications.

\subsection{Potential Outcomes}

The proposed framework is expected to yield improvements in several areas:
\begin{itemize}
    \item \textbf{Test Coverage:} The agent is anticipated to explore a wider variety of navigational paths, potentially uncovering more UI states and identifying hidden defects compared to traditional scripted tests.
    \item \textbf{Defect Detection:} The increase in exploratory actions should result in a higher defect discovery rate, as the agent actively searches for inconsistencies and errors.
    \item \textbf{Reduction in Manual Effort:} The automated generation of BDD test cases can significantly reduce the time and effort required for writing and maintaining test scripts.
    \item \textbf{Adaptive Learning:} Over time, the agent is expected to refine its navigation strategy by learning from past interactions, thereby converging to a more optimal testing policy.
\end{itemize}

\subsection{Evaluation Metrics and Visualization Techniques}

Several evaluation metrics and corresponding visualization techniques are proposed:
\begin{itemize}
    \item \textbf{Learning Curves:}
    \begin{itemize}
        \item \textbf{Reward Trajectory:} Plotting the cumulative or average reward per episode as a line graph to assess convergence behavior and policy improvement.
        \item \textbf{Success Rate Trend:} A line graph or moving average plot illustrating the evolution of the task success rate over time.
    \end{itemize}
    \item \textbf{Test Coverage Visualization:}
    \begin{itemize}
        \item \textbf{Coverage Heatmaps:} Heatmaps to display the frequency of state visits across different website areas, highlighting well-explored and under-explored regions.
        \item \textbf{Bar Charts:} Comparative bar charts showing the number of unique pages or UI elements visited by the RL agent versus those covered by baseline testing methods.
    \end{itemize}
    \item \textbf{Defect Discovery Analysis:}
    \begin{itemize}
        \item \textbf{Scatter Plots:} Plotting defect discovery rates against the number of episodes or steps, to identify episodes with high defect yields.
        \item \textbf{Boxplots:} Displaying the distribution of defects found per episode to facilitate comparisons between different experimental setups.
    \end{itemize}
    \item \textbf{Policy and Trajectory Analysis:}
    \begin{itemize}
        \item \textbf{Trajectory Visualization:} Graphically representing sample state-action sequences and overlaying them on a schematic of the website to better understand navigation paths.
        \item \textbf{Backtracking Frequency:} Plotting the frequency of backtracking events per episode to evaluate how often the agent encounters and recovers from dead-ends.
    \end{itemize}
\end{itemize}

\subsection{Statistical Analysis}

To validate the performance of the approach, the following statistical analyses can be performed:
\begin{itemize}
    \item \textbf{Hypothesis Testing:} Conducting t-tests or similar statistical tests to compare key metrics such as success rates and defect discovery rates between the RL-driven approach and baseline methods.
    \item \textbf{Confidence Intervals:} Calculating confidence intervals for average rewards, test coverage, and episode lengths to assess the reliability of the improvements observed.
\end{itemize}

\subsection{Discussion of Potential Results}

The analysis is expected to address the following:
\begin{enumerate}
    \item \textbf{Coverage Improvement:} The extent to which the RL agent's exploration strategy leads to higher test coverage compared to traditional methods.
    \item \textbf{Defect Detection Efficacy:} Whether the exploratory actions yield a significantly higher rate of defect identification.
    \item \textbf{Learning Dynamics:} Insights into how the learning curve evolves over time and whether the policy converges to an optimal strategy.
    \item \textbf{Trajectory Insights:} An analysis of successful navigation trajectories and their conversion into reusable BDD test cases.
\end{enumerate}

Standard visualization tools such as matplotlib can be used to generate clear and interpretable graphs, providing valuable insights into both the strengths and limitations of the proposed framework.

\section{Conclusion}

This paper has presented a novel framework that integrates an autonomous reinforcement learning (RL) agent with a Behavior-Driven Development (BDD) framework for automated UI testing. By modeling the website as a maze with defined starting points and endpoints, the proposed method enables the agent to dynamically generate and refine test scenarios that align with specific business requirements and user behaviors. A comprehensive system architecture was described, including detailed discussions on state representation, action space definition, reward mechanisms, and exploration strategies utilizing established RL algorithms such as Deep Q-Networks and Policy Gradient methods.

The choice of DQN is justified for its efficiency in discrete action spaces, while Policy Gradient methods provide the necessary flexibility for more complex tasks. The combination of epsilon-greedy exploration and backtracking mechanisms further ensures robust and adaptive navigation of dynamic web interfaces. In addition, the conversion of successful trajectories into human-readable BDD scenarios bridges the gap between automated testing and manual verification.

The proposed framework is expected to improve test coverage, enhance defect detection, and reduce manual effort, thereby transforming the process of software quality assurance. Future work will focus on extending this framework to accommodate more complex web environments and further refining the state representation and reward mechanisms. Moreover, exploring transfer learning and multi-task learning is anticipated to further increase the robustness and generalizability of the RL agent.

\bibliographystyle{unsrt}  
\bibliography{references}

\end{document}